\documentclass[12pt]{article}

\usepackage{enumitem}
\usepackage{graphicx}
\usepackage{float}
\usepackage{cite}
\usepackage{amsfonts}
\usepackage{amssymb}
\usepackage{amsmath}
\usepackage{xcolor}

\def\gtwid{\mathrel{\raise.3ex\hbox{$>$\kern-.75em\lower1ex\hbox{$\sim$}}}}
\def\ltwid{\mathrel{\raise.3ex\hbox{$<$\kern-.75em\lower1ex\hbox{$\sim$}}}}
\def\square{\kern1pt\vbox{\hrule height 1.2pt\hbox{\vrule width 1.2pt\hskip 3pt
  \vbox{\vskip 6pt}\hskip 3pt\vrule width 0.6pt}\hrule height 0.6pt}\kern1pt}

\begin{document}

\begin{titlepage}

\begin{flushright}
CCTP-2025-06 \\
ITCP/2025/06 \\
UFIFT-QG-25-03
\end{flushright}

\vskip 1cm

\begin{center}
{\bf Leading Logarithm Quantum Gravity II}
\end{center}

\vskip 1cm

\begin{center}
S. P. Miao$^{1\star}$, N. C. Tsamis$^{2\dagger}$ and 
R. P. Woodard$^{3\ddagger}$
\end{center}

\vskip 0.5cm

\begin{center}
\it{$^{1}$ Department of Physics, National Cheng Kung University, \\
No. 1 University Road, Tainan City 70101, TAIWAN}
\end{center}

\begin{center}
\it{$^{2}$ Institute of Theoretical Physics \& Computational Physics, \\
Department of Physics, University of Crete, \\
GR-700 13 Heraklion, HELLAS}
\end{center}

\begin{center}
\it{$^{3}$ Department of Physics, University of Florida,\\
Gainesville, FL 32611, UNITED STATES}
\end{center}

\vspace{0cm}

\begin{center}
ABSTRACT
\end{center}
We derive and simplify the gravitational equations that 
apply in accelerating cosmological spacetimes. Solutions 
to these equations should be  tantamount to all order 
re-summations of the perturbative leading logarithms. 
We also discuss possible phenomenological applications 
to cosmological observables.

\begin{flushleft}
\end{flushleft}

\vskip 0.5cm

\begin{flushleft}
$^{\star}$ e-mail: spmiao5@mail.ncku.edu.tw \\
$^{\dagger}$ e-mail: tsamis@physics.uoc.gr \\
$^{\ddagger}$ e-mail: woodard@phys.ufl.edu
\end{flushleft}

\end{titlepage}

\section{Prologue}

Although this paper builds on previous work 
\cite{Miao:2024shs}, the results of which shall 
be used throughout, it is a self-contained study. 
The physical environment is that of early cosmology; 
its geometry is characterized by a scale factor 
$a(t)$ and its two first time derivatives, the 
Hubble parameter $H(t)$ and the 1st slow roll 
parameter $\epsilon(t)$:
\footnote{It is more often than not convenient to 
employ conformal instead of co-moving coordinates:
$ds^2 \!=\! 
-dt^2 + a^2(t) \, d{\mathbf{x}} \cdot d{\mathbf{x}} 
= 
a^2(\eta) \big[\! -d\eta^2 + a^2(\eta) \, 
d{\mathbf{x}} \cdot d{\mathbf{x}} \big]$, with
$t$ the co-moving time and $\eta$ the conformal
time.}
\begin{equation}
ds^2 = - dt^2 + a^2(t) \, d{\mathbf{x}} \cdot d{\mathbf{x}} 
\qquad \Longrightarrow \qquad 
H(t) \equiv \frac{\dot{a}}{a} 
\quad, \quad 
\epsilon(t) \equiv - \frac{\dot{H}}{H^2} 
\; . \label{geometry}
\end{equation}
In the very early universe, primordial inflation is an 
era of accelerated expansion ($H > 0$ with $0 \leq \epsilon 
< 1$). During this era virtual particles are ripped out 
of the vacuum \cite{Schrodinger:1939} and the phenomenon 
is largest for particles such as massless, minimally coupled 
(MMC) scalars and gravitons, because they are both massless 
and not conformally invariant \cite{Lifshitz:1945du,
Grishchuk:1974ny}. This particle production is thought to 
be the physical mechanism causing the primordial tensor
\cite{Starobinsky:1979ty} and scalar \cite{Mukhanov:1981xt} 
power spectra.

As inflation progresses more and more quanta are created
so that correlators which involve interacting MMC scalars 
and gravitons often show secular growth in the form of 
powers of $\ln[a(t)]$ \cite{Woodard:2025cez}. 
\footnote{Examples of this particular secular growth behaviour 
can be found in citations [14,16,24,31,32,38,44,72,73,77,78] 
therein.}
Terms with the maximum power of $\ln(a)$ at a given perturbative
order are known as {\it leading logarithm} (LLOG) contributions; 
those with fewer factors of $\ln(a)$ are known as {\it subleading 
logarithm} contributions.

There are two sources of secular logarithms in inflationary
spacetimes:
\\
{\it (i)} the ``tail'' logarithmic term in the graviton 
propagator; very schematically:
\begin{equation}
i \Delta \sim \tfrac{1}{x^2} - \ln(x^2)
\; , \label{TailLogs} 
\end{equation}
{\it (ii)} the renormalization procedure via dimensional 
regularization; very briefly the primitive divergences 
do not contain $D$-dependent scale factors while the 
counterterms do and hence there is an incomplete cancellation:
\begin{equation}
\tfrac{H^{D-4}}{D-4} - \tfrac{(\mu a)^{D-4}}{D-4} 
= 
- \ln\left( \tfrac{\mu a}{H} \right) + \mathcal{O}(D-4)
\; . \label{UVlogs}
\end{equation}
In this paper we shall be concerned with the {\it first}
source and postpone the study of the second source for
the near future.

In a long period of inflation factors of $\ln[a(t)]$ can grow 
large enough to overwhelm even the smallest coupling constant. 
The dimensionless coupling constant of pure quantum gravity is 
$G H^2$ and at some time the secular increase by powers of 
$\ln[a(t)]$ will overwhelm $G H^2$ causing perturbation theory 
to break. Physically, the most interesting particle to study 
is clearly the graviton; can the universally attractive nature 
of the gravitational interaction alter cosmological parameters, 
kinematical parameters and long-range forces? A preliminary 
study of this, albeit with ``semi-primitive'' for the intended 
purpose quantum field theoretic tools, {\it indicated} 
a positive answer \cite{Tsamis:1994ca,Tsamis:2011ep}.

Ultimately, we must try to re-sum the leading logarithms 
of pure gravity and hopefully obtain the late time limits 
of cosmological correlators. While it was easy to state what 
is needed - a re-summation technique for the leading logarithms 
- the realization of non-perturbative techniques is invariably
very hard. The re-summation technique we shall consider and 
in our mind has been adequately developed, is the stochastic 
technique pioneered by the late Alexei Starobinsky 
\cite{Starobinsky:1986fx}. The main objective of this paper 
is to derive the LLOG equations of motion for pure gravity 
whose solution would re-sum the leading logarithms and 
hopefully give us interesting physical results regarding 
the dynamics of inflation.

This paper consists of four Sections, of which this 
Prologue was the first, and one Appendix. 
In Section 2 we summarize the results of the first 
paper \cite{Miao:2024shs} commencing from pure gravity 
in a constant background (sub-section 2.1), to the 
space+time decomposition (sub-section 2.2), to the 
relevant Feynman rules (sub-section 2.3), to the 
optimal decomposition of the total Lagrangian 
(sub-section 2.4). Section 2 concludes with the 
resulting gravitational equations (sub-section 2.5). 
Section 3 starts by summarizing the LLOG re-summation 
techniques of \cite{Miao:2024shs} (sub-section 3.1), 
and then proceeds to the main results of this work by 
exhibiting the desired LLOG equations that pure quantum 
gravity implies (sub-sections 3.2-3.4). 
Section 4 is the Epilogue where we discuss the physical 
implications and prospects. 
Finally some useful mathematical facts are catalogued 
in the Appendix.

\newpage

\section{Quantum Gravity}

Pure gravity in $D$-dimensions is defined by the Lagrangian:
\footnote{Hellenic indices take on spacetime values
while Latin indices take on space values. Our metric
tensor $g_{\mu\nu}$ has spacelike signature
$( - \, + \, + \, +)$ and our curvature tensor equals
$R^{\alpha}_{~ \beta \mu \nu} \equiv
\Gamma^{\alpha}_{~ \nu \beta , \mu} +
\Gamma^{\alpha}_{~ \mu \rho} \,
\Gamma^{\rho}_{~ \nu \beta} -
(\mu \leftrightarrow \nu)$.}
\begin{equation}
{\mathcal L}_{inv} =
\frac{1}{\kappa^2} \big[ R \sqrt{-g} - (D-2)(D-1) H^2 \sqrt{-g} \, \big]
\; , \label{Linv}
\end{equation}
and possesses two dimensionful parameters, Newton's
constant $G$ and the cosmological constant $\Lambda$:
\footnote{Notice that even for a cosmological mass scale 
$\, M \!\sim\! 10^{18} GeV \,$ close to the Planck scale 
$M_{\rm Pl}$, 
the dimensionless coupling constant is very small: 
$G \Lambda \!=\! \tfrac{M^4}{M^4_{\rm Pl}} \!\sim\! 10^{-4}$.}
\begin{equation}
\kappa^2 \equiv 16 \pi G
\quad , \quad
\Lambda \equiv (D-1) H^2
\; . \label{parameters}
\end{equation}
The resulting $D$-dimensional equations of motion are:
\begin{equation}
R_{\mu\nu} - \frac12 R \, g_{\mu\nu} 
+ \frac12 (D-2)(D-1) H^2 g_{\mu\nu} = 0
\; . \label{eom}
\end{equation}
In terms of the full metric $g_{\mu\nu}$, the conformally
rescaled metric ${\widetilde g}_{\mu\nu}$ and the graviton 
field $h_{\mu\nu}$ are defined thusly:
\begin{equation}
g_{\mu\nu} \equiv 
a^2 {\widetilde g}_{\mu\nu} \equiv
a^2 \big[ \eta_{\mu\nu} + \kappa h_{\mu\nu} \big] 
\; . \label{metrics}
\end{equation}

In view of (\ref{metrics}) the Lagrangian (\ref{Linv})
can be written as follows \cite{Tsamis:1992xa}:
\begin{eqnarray}
{\mathcal L}_{inv} &\!\!\!=\!\!\!&
a^{D-2} \sqrt{-{\widetilde g}} \, {\widetilde g}^{\alpha\beta}
{\widetilde g}^{\rho\sigma} {\widetilde g}^{\mu\nu} 
\Big\{ 
\tfrac12 h_{\alpha\rho, \mu} h_{\nu\sigma, \beta} 
- \tfrac12 h_{\alpha\beta, \rho} h_{\sigma\mu, \nu}
+ \tfrac14 h_{\alpha\beta, \rho} h_{\mu\nu, \sigma}
\nonumber \\
& \mbox{} & 
- \tfrac14 h_{\alpha\rho, \mu} h_{\beta\sigma, \nu}
\Big\}  
+ (\tfrac{D}{2} - 1) a^{D-1} H 
\sqrt{-{\widetilde g}} \, {\widetilde g}^{\rho\sigma}
{\widetilde g}^{\mu\nu}
h_{\rho\sigma, \mu} h_{\nu 0}
\; , \label{Linv2}
\end{eqnarray}
which is the form of $\mathcal{L}_{inv}$ we shall use
thereafter.

\subsection{LLOG history}

The simplest version of Starobinsky's stochastic formalism
\cite{Starobinsky:1986fx} can be proven to capture the leading
logarithms of MMC scalar potential models \cite{Tsamis:2005hd}.
Generalizing it to include derivative interactions like those
of gravity was a long and tedious struggle, which this paper 
should hopefully complete. The first step was the realization 
that MMC scalars coupled to ``passive'' fields such as fermions 
\cite{Miao:2006pn} and photons \cite{Prokopec:2007ak}, can be 
reduced to scalar potential models by simply integrating
out the passive fields in a constant MMC scalar background. 
\footnote{``Passive'' fields are defined as those whose 
propagator does not possess a logarithmic tail term 
(\ref{TailLogs}).} 
This induces the standard kind of effective potential based 
on altered masses. It was next shown that non-linear sigma 
models can be treated by integrating out differentiated scalars 
in a constant scalar background from the equations of motion 
\cite{Miao:2021gic}. This induces a new sort of effective 
potential based on altered field strengths, rather than altered 
masses. Finally, it was shown that MMC scalar corrections to 
gravity can be treated by integrating out differentiated scalars 
from the stress tensor in the presence of a constant graviton 
background \cite{Miao:2024nsz}. This induces a new sort of 
effective potential (really an induced stress tensor) based on 
altering the Hubble parameter. The previous paper in this
series \cite{Miao:2024shs} worked out how to modify the gauge 
fixing functional so that differentiated graviton fields can be
integrated out in the presence of a constant graviton background.

\subsection{Gravity in a constant background}

The standard paradigm of a primordial inflationary
spacetime is the de Sitter (dS) maximally symmetric
geometry:
\footnote{The Feynman rules for dS have been summarized
in \cite{Miao:2024shs}.}
\begin{equation}
{\widetilde g}^{dS}_{\mu\nu} = \eta_{\mu\nu}
\; . \label{dS}
\end{equation}
The quantum physics of de Sitter spacetime has a long, 
and sometimes controversial, history to this date,
e.g. \cite{Allen:1985wd, Antoniadis:1986sb, Allen:1986tt, 
Allen:1987tz, Floratos:1987ek, Burgess:2010dd, 
Marolf:2012kh, Anderson:2013ila, Anninos:2014lwa, 
Dvali:2014gua, Burgess:2015ajz, Brandenberger:2018fdd, 
Cable:2023gdz, Anninos:2024fty, Bhowmick:2025mxh, 
Kamenshchik:2025ses, Sahni:2025jdr, Li:2025azq}.

To accomodate the LLOG approximation in pure gravity 
(\ref{Linv}) we shall need to do so in the presence 
of a constant graviton background \cite{Miao:2024shs}. 
For a background with constant $H$ and arbitrary 
$\widetilde{g}_{\mu\nu}$:
\begin{equation}
g_{\mu\nu}(x) \equiv 
a^2 \, \widetilde{g}_{\mu\nu}(x) \equiv 
a^2 \big[ \eta_{\mu\nu} + \kappa h_{\mu\nu}(x) \big] 
\quad , \quad 
a = -(H \eta)^{-1}
\; , \label{background}
\end{equation}
the curvature tensor when $\widetilde{g}_{\mu\nu}$ 
is constant equals:
\begin{equation}
R^{\rho}_{~\sigma\mu\nu} \Big\vert_{\widetilde{g}_{\mu\nu} = c}
=
-H^2 \, \widetilde{g}^{00} \bigl(
\delta^{\rho}_{~\mu} \, g_{\sigma \nu}
- \delta^{\rho}_{~\nu} \, g_{\sigma\mu} \bigr)
\; , \label{RiemannConstant}
\end{equation}
which is a de Sitter geometry but with a {\it different} 
cosmological constant:
\begin{equation}
\widetilde{g}_{\mu\nu , \rho} = 0
\quad \Longrightarrow \quad
H^2
\; \longrightarrow \;
-\widetilde{g}^{00} H^2
\; . \label{constantHshift}
\end{equation}

However, because gravitons have tensor indices, the gauge
fixing procedure must be extended to accomodate the arbitrary
constant graviton background. The most efficient way to extend
the graviton Feynman rules from de Sitter to any spacetime
such that $\, {\widetilde g}_{\mu\nu, \rho} = \,0$, starts
with the (D-1)+1 decomposition.

\subsection{The (D-1)+1 decomposition}

As described in \cite{Miao:2024shs}, it is most desirable 
to employ the ADM decomposition of the conformally rescaled 
metric \cite{Arnowitt:1962hi}:
\footnote{In (\ref{3+1lower}-\ref{3+1upperB}) $N$ 
is the lapse function, $N^i$ is the shift function, 
and $\gamma_{ij}$ is the spatial metric. Furthermore, 
${\overline \gamma}^{\mu\nu}$ is seen to be the 
``spatial part'' and $u_{\mu}$ the ``temporal part''.} 
\begin{eqnarray}
{\widetilde g}_{\mu\nu} 
&\!\!\!\! = \!\!\!\!&
\begin{pmatrix}
-N^2 \!+\! \gamma_{kl} N^k N^l \;&\; -\gamma_{jl} N^l \\
\\
-\gamma_{ik} N^k & \gamma_{ij} \\
\end{pmatrix}
\label{3+1lower} \\
&\!\!\!\! = \!\!\!\!&
\begin{pmatrix}
\gamma_{kl} N^k N^l \;&\; -\gamma_{jl} N^l \!&\! \\
\\
-\gamma_{ik} N^k \;&\; \gamma_{ij} \\
\end{pmatrix}
-
\begin{pmatrix}
-N \\
\\
0 \\
\end{pmatrix}
_{\!\!\!\mu}
\begin{pmatrix}
-N \\
\\
0 \\
\end{pmatrix}
_{\!\!\!\nu} 
\equiv
{\overline \gamma}_{\mu\nu} \!\!- u_{\mu} u_{\nu}
\; , \qquad \label{3+1lowerB}
\end{eqnarray}
which implies the following form for its inverse: 
\begin{eqnarray}
{\widetilde g}^{\mu\nu}
&\!\!\!\! = \!\!\!\!&
\begin{pmatrix}
-\frac{1}{N^2} & -\frac{N^j}{N^2} \\
\\
-\frac{N^i}{N^2} \;&\; \gamma^{ij} \!-\! \frac{N^i N^j}{N^2} \\
\end{pmatrix}
\label{3+1upper} \\
&\!\!\!\! = \!\!\!\!&
\begin{pmatrix}
0 & 0 \\
\\
0 \;&\; \gamma^{ij} \\
\end{pmatrix}
-
\begin{pmatrix}
\frac{1}{N} \\
\\
\frac{N^i}{N} \\
\end{pmatrix}
^{\!\!\!\mu}
\begin{pmatrix}
\frac{1}{N} \\
\\
\frac{N^j}{N} \\
\end{pmatrix}
^{\!\!\!\nu} 
\equiv
{\overline \gamma}^{\mu\nu} \!\!- u^{\mu} u^{\nu}
\; . \qquad \label{3+1upperB}
\end{eqnarray}

\subsection{The Feynman rules}

We simply display the results from \cite{Miao:2024shs} 
concerning:
\footnote{Some useful properties of the propagators 
and their derivatives can be found in the Appendix.}
\\ [5pt]
$\bullet$ {\it The gauge fixing sector}
\\
The extended gauge fixing Lagrangian term equals:
\begin{equation}
{\widetilde{\mathcal L}}_{GF} =
- \tfrac12 a^{D-2} \sqrt{-{\widetilde g}} \, 
{\widetilde g}^{\mu\nu} \, {\widetilde F}_{\mu} \,
{\widetilde F}_{\nu}
\; , \label{Lgf1} \\
\end{equation}
and the extended gauge condition is:
\begin{equation}
{\widetilde F}_{\mu} =
{\widetilde g}^{\rho\sigma} \, 
\Big[ h_{\mu\rho, \sigma} - \tfrac12 h_{\rho\sigma, \mu} \,
- (D-2) a {\widetilde H} h_{\mu\rho} u_{\sigma} \Big]
\quad , \quad 
{\widetilde H} \equiv \frac{H}{N}
\; . \qquad \label{F} 
\end{equation}
To arrive at the desired form for the gauge fixing 
Lagrangian term, we substitute (\ref{F}) 
into (\ref{Lgf1}):
\begin{eqnarray}
{\widetilde{\mathcal L}}_{GF}
&\!\!\!=\!\!\!&
a^{D-2} \sqrt{-{\widetilde g}} \, {\widetilde g}^{\alpha\beta}
{\widetilde g}^{\rho\sigma} {\widetilde g}^{\mu\nu} 
\Big\{ 
- \tfrac12 h_{\mu\rho, \sigma} h_{\nu\alpha, \beta} 
+ \tfrac12 h_{\mu\rho, \sigma} h_{\alpha\beta, \nu}
- \tfrac18 h_{\rho\sigma, \mu} h_{\alpha\beta, \nu}
\qquad \nonumber \\
& \mbox{} & 
+ (D-2) a {\widetilde H} h_{\mu\rho, \sigma} h_{\nu\alpha} u_{\beta}
- (\tfrac{D}{2}-1) a {\widetilde H} h_{\rho\sigma, \mu} h_{\nu\alpha} 
u_{\beta}
\nonumber \\
& \mbox{} & 
- \tfrac12 (D-2)^2 a^2 {\widetilde H}^2 h_{\mu\rho} h_{\nu\alpha} 
u_{\beta} u_{\sigma}
\Big\} 
\label{Lgf2}
\end{eqnarray}
$\bullet$ {\it The graviton sector} 
\\
The graviton propagator is:
\begin{equation}
i [\mbox{}_{\mu\nu} \widetilde{\Delta}_{\rho\sigma}](x;x') 
= 
\sum_{I=A,B,C} [\mbox{}_{\mu\nu} \widetilde{T}^I_{~\rho\sigma}] 
\!\times
i \widetilde{\Delta}_{I}(x;x') 
\; , \label{gravprop} 
\end{equation}
where $i \widetilde{\Delta}_{A}(x;x')$ is the massless minimally 
coupled scalar propagator in dS with Hubble parameter $\widetilde{H}$, 
$i \widetilde{\Delta}_{B}(x;x')$ is the massive scalar propagator in 
dS for $m^2 = (D-2) \widetilde{H}^2$, 
and $i \widetilde{\Delta}_{C}(x;x')$ is the massive scalar propagator 
in dS for $m^2 = 2(D-3) \widetilde{H}^2$. 
The tensor factors, as per usual in our class of gauges, have the 
remarkable property of being spacetime {\it constants} and hence 
the resulting propagator has excellent perturbative calculability:
\footnote{Furthermore, in $D=4$ the three scalar propagators
$i \widetilde{\Delta}_{I}(x;x')$ {\it only} have one or two terms.} 
\begin{eqnarray}
[\mbox{}_{\mu\nu} \widetilde{T}^{A}_{~\rho\sigma}] 
&\!\!\! = \!\!\!& 
2 \overline{\gamma}_{\mu (\rho} \overline{\gamma}_{\sigma) \nu} 
- \tfrac{2}{D-3} \overline{\gamma}_{\mu\nu} \overline{\gamma}_{\rho\sigma} 
\; , \label{T_A} \\
{[} \mbox{}_{\mu\nu} \widetilde{T}^{B}_{~\rho\sigma}] 
&\!\!\! = \!\!\!& 
- 4 u_{(\mu} \overline{\gamma}_{\nu) (\rho} u_{\sigma)} 
\; , \label{T_B} \\
{[} \mbox{}_{\mu\nu} \widetilde{T}^{C}_{~\rho\sigma}] 
&\!\!\! = \!\!\!&
\tfrac{2}{(D-3)(D-2)}
\big[ (D-3) u_{\mu} u_{\nu} + \overline{\gamma}_{\mu\nu} \big] 
\big[ (D-3) u_{\rho} u_{\sigma} + \overline{\gamma}_{\rho\sigma} \big] 
\; . \label{T_C}
\end{eqnarray}
$\bullet$ {\it The ghost sector}
\\
The ghost and antighost fermionic fields $c_{\mu}$ and 
${\overline c}_{\mu}$ contribute thusly:
\begin{equation}
{\widetilde{\mathcal L}}_{gh}
=
- a^{D-2} \sqrt{-{\widetilde g}} \, {\widetilde g}^{\mu\nu}
{\overline c}_{\mu} \, \delta F_{\nu}
\; , \label{Lgh}
\end{equation}
where in the infinitesimal variation $\delta F_{\nu}$ 
of the gauge fixing condition (\ref{F}) the variation 
parameter is the ghost field $c$. 

Perhaps a more appropriate re-writing of (\ref{Lgh}) is:
\footnote{We thank B. Yesilyurt for pointing out a mistake 
in an earlier version \cite{Miao:2024shs}.}
\begin{equation}
\widetilde{\mathcal{L}}_{\rm ghost} 
= 
a^{D-2} \sqrt{-\widetilde{g}} \, \overline{c}^{\alpha} 
\, \widetilde{\mathcal{G}}_{\alpha\beta} \, c^{\beta}.   
\; , \label{Lgh2}
\end{equation}
where the differential operator 
$\widetilde{\mathcal{G}}_{\alpha\beta}$ equals:
\begin{eqnarray}
\widetilde{\mathcal{G}}_{\alpha\beta} 
& \!\!\!=\!\!\! & 
\widetilde{g}_{\alpha\beta} \widetilde{g}^{\rho\sigma} 
\partial_{\rho} \partial_{\sigma} 
- {\scriptstyle (D-2)} \, a \widetilde{H} \eta_{\alpha\beta} \, u^{\rho} 
\partial_{\rho} 
+ {\scriptstyle (D-2)} \, a^2 \widetilde{H}^2 u_{\alpha} u_{\beta} 
\nonumber \\
& & \hspace{-0.9cm}
+ \, \widetilde{g}^{\rho\sigma} \kappa ( h_{\beta\rho , \sigma} 
\!-\! \tfrac12 h_{\rho\sigma , \beta} ) \partial_{\alpha} 
+ \big[ \, \widetilde{g}^{\rho\sigma} \kappa ( h_{\alpha\rho , \sigma} 
\!-\! \tfrac12 h_{\rho\sigma , \alpha} ) \big]_{,\beta} 
\nonumber \\
& & \hspace{-0.9cm}
+ {\scriptstyle (D-2)} \, a \widetilde{H} \eta_{\alpha\rho} 
u_{\sigma} \widetilde{g}^{\rho\sigma}_{~~,\beta}  
+ {\scriptstyle (D-2)} \, a \widetilde{H} \kappa h_{\alpha\rho} 
\widetilde{g}^{\rho\sigma} u_{\beta} \partial_{\sigma} 
- {\scriptstyle 2 (D-2)} \, a^2 \widetilde{H}^2 \kappa h_{\alpha\rho} 
u^{\rho} u_{\beta} 
\; . \qquad \label{Gtilde} 
\end{eqnarray}
The ghost propagator takes the form: 
\begin{equation}
i[\mbox{}_{\mu} \widetilde{\Delta}_{\rho}](x;x') 
= 
\overline{\gamma}_{\mu\rho} \!\times i \widetilde{\Delta}_A(x;x') 
- u_{\mu} u_{\rho} \!\times i \widetilde{\Delta}_{B}(x;x') 
\; . \label{ghostprop}
\end{equation}

\subsection{The optimal decomposition}

The total Lagrangian is the usual sum of three terms: 
the invariant part $\mathcal{L}_{inv}$, the gauge fixing
part $\mathcal{\widetilde L}_{GF}$, and the ghost part
$\mathcal{\widetilde L}_{gh}$:
\begin{equation}
\mathcal{L}_{tot} =
\mathcal{L}_{inv} + \mathcal{\widetilde L}_{GF} +
\mathcal{\widetilde L}_{gh}
\; . \label{Ltot}
\end{equation}
It will turn out that for our purposes from now onwards 
it is mathematically correct to take the $D=4$ limit.
\footnote{In the next Section we shall explain why the 
$D=4$ limit suffices.}
Furthermore, it is preferable to add the contributions 
of $\mathcal{L}_{inv}$ (\ref{Linv2}) and 
$\mathcal{\widetilde L}_{GF}$ (\ref{Lgf2}) because some 
of their terms simplify against each other. The six terms 
comprising $\, \mathcal{L}_{inv} \!+\! 
\mathcal{\widetilde L}_{GF} \,$ can be grouped into three 
convenient parts:
\begin{equation}
\mathcal{L}_{inv} + \mathcal{\widetilde L}_{GF}
\; \equiv \;
\mathcal{L}_{1+2+3} + \mathcal{L}_{4+5} + \mathcal{L}_6
\; , \label{Linv+Lgf}
\end{equation}
where these three convenient parts in $D=4$ equal:
\begin{eqnarray}
\mathcal{L}_{1+2+3} 
&\!\!\! = \!\!\!&
a^2 \sqrt{-{\widetilde g}} \, {\widetilde g}^{\alpha\beta}
{\widetilde g}^{\rho\sigma} {\widetilde g}^{\gamma\delta} 
\nonumber \\
& \mbox{} & \hspace{0.3cm}
\times \,\Big\{ \!
- \tfrac14 h_{\alpha\rho, \gamma} h_{\beta\sigma, \delta} 
+ \tfrac18 h_{\alpha\beta, \gamma} h_{\rho\sigma, \delta}
+ a^2 {\widetilde H}^2 h_{\gamma\rho} u_{\sigma} 
  h_{\delta\alpha} u_{\beta} \Big\}
\; , \qquad \label{L1+2+3} \\
\mathcal{L}_{4+5} 
&\!\!\! = \!\!\!&
\sqrt{-{\widetilde g}} \, {\widetilde g}^{\alpha\beta}
{\widetilde g}^{\rho\sigma} {\widetilde g}^{\gamma\delta}
\partial_{\beta} \Big[ 
- \tfrac12 \partial_{\sigma} \big( a^2 h_{\gamma\rho} 
  h_{\delta\alpha} \big)
+ a^2 h_{\gamma\rho} h_{\delta\alpha, \sigma} \Big]
\; , \label{L4+5} \\
\mathcal{L}_6
&\!\!\! = \!\!\!&
\kappa a^3 H \sqrt{-{\widetilde g}} \, {\widetilde g}^{\alpha\beta}
{\widetilde g}^{\rho\sigma} {\widetilde g}^{\gamma\delta}
h_{\rho\sigma, \gamma} h_{\delta\alpha} h_{0\beta}
\; . \label{L6} 
\end{eqnarray}
Finally the ghost contribution $\mathcal{\widetilde L}_{gh}$
of interest is given by (\ref{Lgh2}) after we set $D=4$ in 
(\ref{Gtilde}).

\subsection{The resulting equations}
The starting point of our analysis are the $D=4$ gravitational
field equations that follow from $\mathcal{L}_{tot}$ when we 
consider the three convenient parts (\ref{L1+2+3}-\ref{L6}):
\begin{eqnarray}
\frac{\delta S_{1+2+3}}{\delta h_{\mu\nu}} 
& \!\!\!=\!\!\! &
\partial_{\alpha} \Bigl\{ 
a^2 \sqrt{-\widetilde{g}} \, \widetilde{g}^{\alpha\beta} 
( \tfrac12 \widetilde{g}^{\mu \rho} \widetilde{g}^{\nu \sigma} 
\!\!-\! \tfrac14 \widetilde{g}^{\mu\nu} \widetilde{g}^{\rho\sigma} ) 
h_{\rho\sigma , \beta} \Bigr\} 
\hspace{4.1cm} \nonumber \\
& &
+ 2 a^4 \sqrt{-\widetilde{g}} \, \widetilde{H}^2 
u^{(\mu} \widetilde{g}^{\nu) \alpha} u^{\beta} h_{\alpha\beta}
\nonumber \\
& & 
+ a^2 \tfrac{\partial}{\partial h_{\mu\nu}} \Bigl\{
\sqrt{-\widetilde{g}} \, \widetilde{g}^{\alpha\beta} 
( -\tfrac14 \widetilde{g}^{\rho\gamma} \widetilde{g}^{\sigma\delta} 
+ \tfrac18 \widetilde{g}^{\rho\sigma} \widetilde{g}^{\gamma\delta}) \Bigr\} 
h_{\gamma\delta , \alpha} h_{\rho\sigma , \beta} 
\nonumber \\
& & 
+ a^4 \widetilde{H}^2 \tfrac{\partial}{\partial h_{\mu\nu}} 
\Bigl\{ \sqrt{-\widetilde{g}} \, \widetilde{g}^{\gamma\alpha} 
\widetilde{g}^{\beta\rho} \widetilde{g}^{\sigma\delta} \Bigr\} 
u_{\gamma} u_{\delta} h_{\alpha\beta} h_{\rho\sigma} 
\; , \label{L1+2+3eom}
\end{eqnarray}
\begin{eqnarray}
\frac{\delta S_{4+5}}{\delta h_{\mu\nu}} 
& \!\!\!=\!\!\! & 
\tfrac{\partial}{\partial h_{\mu\nu}} \Bigl\{ 
\sqrt{-\widetilde{g}} \, \widetilde{g}^{\alpha\beta} 
\widetilde{g}^{\gamma\delta} \widetilde{g}^{\rho\sigma} \Bigr\}
\hspace{7.7cm} \nonumber \\
& & \hspace{1.3cm}
\times \Bigl\{ -3 a^4 \widetilde{H}^2 u_{\beta} u_{\sigma} 
h_{\gamma\rho} h_{\delta\alpha} 
+ \tfrac{a^2}{2} ( h_{\gamma\rho , \beta} h_{\delta\alpha , \sigma}  
- h_{\gamma\rho , \sigma} h_{\delta\alpha , \beta} ) \Bigr\} 
\nonumber \\
& &
+ a^3 \sqrt{-\widetilde{g}} \, \widetilde{H} 
( 2 \widetilde{g}^{\alpha(\mu} u^{\nu)} 
\!\!-\! u^{\alpha} \widetilde{g}^{\mu\nu} )
h_{\alpha\beta} \, \widetilde{g}^{\beta \rho}_{~~,\rho} 
\nonumber \\
& & 
- 2 a^3 (\widetilde{H} \sqrt{-\widetilde{g}} \, u^{\alpha})_{,\rho} 
\, h_{\alpha\beta} \, \widetilde{g}^{\beta (\mu} 
\widetilde{g}^{\nu) \rho} 
\nonumber \\
& &
+ a^2( h_{\gamma\rho , \sigma} \partial_{\beta} 
\!-\! h_{\gamma\rho , \beta} \partial_{\sigma} ) 
(\sqrt{-\widetilde{g}} \, \widetilde{g}^{\rho\sigma}
\widetilde{g}^{\beta (\mu} \widetilde{g}^{\nu ) \gamma}) 
\; , \label{L4+5eom}
\end{eqnarray}
\begin{eqnarray}
\frac{\delta S_{6}}{\delta h_{\mu\nu}}  
& \!\!\!=\!\!\! & 
-\kappa H \sqrt{-\widetilde{g}} \, \widetilde{g}^{\mu\nu} 
\partial_{\gamma} ( a^3 \widetilde{g}^{\alpha\beta} 
\widetilde{g}^{\gamma\delta} h_{\delta \alpha} h_{\beta 0} )
\hspace{5.9cm} \nonumber \\
& &
+ 2 a^3 H \tfrac{\partial}{\partial h_{\mu\nu}} \Bigl\{
\widetilde{g}^{\alpha\beta} \widetilde{g}^{\gamma\delta} 
h_{\alpha\delta} h_{\beta 0} \Bigr\} 
\partial_{\gamma} \sqrt{-\widetilde{g}} 
\; , \label{L6eom}
\end{eqnarray}
as well as the $D=4$ limit of the ghost Lagrangian (\ref{Lgh2})
that gives the most complicated equation of motion by far: 
\begin{eqnarray}
\frac{\delta \widetilde{S}_{\rm ghost}}{\delta h_{\mu\nu}} 
& \!\!\!=\!\!\! & 
\tfrac{\kappa}{2} a^2 \sqrt{-\widetilde{g}} \, \widetilde{g}^{\mu\nu} \, 
\overline{c}^{\alpha} \widetilde{\mathcal{D}}_{\alpha\beta} \, c^{\beta} 
+ \kappa a^2 \sqrt{-\widetilde{g}} \, \overline{c}^{\alpha} \Bigl\{ 
( \delta^{\mu}_{\alpha} \delta^{\nu}_{\beta} \, \widetilde{g}^{\rho\sigma} 
\!-\! \widetilde{g}_{\alpha\beta} \widetilde{g}^{\rho\mu} 
\widetilde{g}^{\sigma\nu} ) \partial_{\rho} \partial_{\sigma}
\nonumber \\
& & \hspace{-0.7cm}
+ 2 a \widetilde{H} \eta_{\alpha\beta} \, u^{(\mu} \widetilde{g}^{\nu) \rho} 
\partial_{\rho} 
\!-\! \widetilde{g}^{\rho\mu} \widetilde{g}^{\sigma\nu} \kappa 
( h_{\beta \rho , \sigma} \!-\! \tfrac12 h_{\rho\sigma , \beta} ) 
\partial_{\alpha} 
\!+\! 2 a \widetilde{H} \eta_{\alpha\rho} \, \widetilde{g}^{\rho (\mu} 
\widetilde{g}^{\nu) \sigma} ) u_{\beta} \partial_{\sigma} 
\nonumber \\
& & 
- 4 a^2 \! \widetilde{H}^2 \eta_{\alpha\rho} \,
\widetilde{g}^{\rho (\mu} u^{\nu)} u_{\beta} \Bigr\} \, c^{\beta}
\nonumber \\
& & \hspace{-0.7cm}
- \kappa \, \partial_{\rho} \Bigl\{a^2 \sqrt{-\widetilde{g}} \, 
\overline{c}^{\alpha} \widetilde{g}^{\rho (\mu} c^{\nu)}_{~~,\alpha} \Bigr\} 
+ \tfrac{\kappa}{2} \, \partial_{\beta} \Bigl\{ 
a^2 \sqrt{-\widetilde{g}} \, \widetilde{g}^{\mu\nu} \, 
\overline{c}^{\alpha} c^{\beta}_{~,\alpha} \Bigr\} 
\nonumber \\
& & \hspace{-0.7cm}  
+ \, \widetilde{g}^{\rho (\mu} \widetilde{g}^{\nu) \sigma} 
\kappa^2 ( h_{\alpha\rho , \sigma} - \tfrac12 h_{\rho\sigma , \alpha} ) \,
\partial_{\beta} \Bigl\{ a^2 \sqrt{-\widetilde{g}} \, 
\overline{c}^{\alpha} c^{\beta} \Bigr\}
\nonumber \\
& & \hspace{-0.7cm}
+ \kappa \, \partial_{\rho} \Bigl\{ 
\widetilde{g}^{\rho (\mu} \partial_{\beta} 
( a^2 \sqrt{-\widetilde{g}} \, \overline{c}^{\nu)} c^{\beta} ) \Bigr\}
- \tfrac{\kappa}{2} \, \partial_{\alpha} \Bigl\{ 
\widetilde{g}^{\mu\nu} \partial_{\beta} ( a^2 \sqrt{-\widetilde{g}} \, 
\overline{c}^{\alpha} c^{\beta}) \Bigr\} 
\nonumber \\
& & \hspace{-0.7cm}
+ 2 \, \widetilde{g}^{\rho (\mu} \widetilde{g}^{\nu) \sigma} \partial_{\beta}
\Bigl\{ a^3 \sqrt{-\widetilde{g}} \, \kappa \widetilde{H} \eta_{\alpha\rho} 
u_{\sigma} \, \overline{c}^{\alpha} c^{\beta} \Bigr\} 
\; , \label{Lgheom}
\end{eqnarray}
where $\widetilde{\mathcal{D}}_{\alpha\beta}
= \widetilde{\mathcal{G}}_{\alpha\beta}$ in $D=4$.

\section{The LLOG gravitational equations}

The equations we wish to obtain are derived by applying 
the LLOG rules for theories with derivative interactions 
(see \cite{Miao:2024shs} section 2.2) to the gravitational 
equations (\ref{L1+2+3eom}-\ref{Lgheom}). These rules
require a series of straightforward and usually quite
cumbersome manipulations but in the end the desired 
equations will assume a rather simple form which we 
present in this section.

\subsection{The LLOG rules for pure gravity}

To begin with, pure gravity is a theory with derivative 
interactions and the relevant rules are given in section
2.2 of \cite{Miao:2024shs}:
\begin{equation}
\frac{\delta S[h]}{\delta h_{\mu\nu}} \Big\vert_{LLOG} 
\equiv
\frac{\delta S[h]}{\delta h_{\mu\nu}} \Big\vert_{stoch}
\! +
\frac{\delta S[h]}{\delta h_{\mu\nu}} \Big\vert_{ind} 
\; \big( \! +
\frac{\delta S_{matter}}{\delta h_{\mu\nu}} \big)
 = 0
\; . \label{QGrule}
\end{equation}
In the above expression (\ref{QGrule}):
\\ [3pt]
$\bullet \;$ $\tfrac{\delta S[h]}{\delta h_{\mu\nu}} \vert_{LLOG}$ 
represents the desired equation of motion capturing the leading 
logarithms to all orders, and is derived from the full Heisenberg 
equation of motion $\tfrac{\delta S[h]}{\delta h_{\mu\nu}}$ by 
adding the three contributions of the RHS.
\\ [3pt]
$\bullet \;$ The first contribution 
$\tfrac{\delta S[h]}{\delta h{\mu\nu}} \vert_{stoch}$ 
represents the {\it ``stochastic reduction''} of the derivative
terms in the Heisenberg field equation 
$\tfrac{\delta S[h]}{\delta h_{\mu\nu}}$.
\footnote{An enormous advantage of the method becomes 
apparent because we must deal with a classical stochastic 
equation instead of the Heisenberg field equations of an 
interacting QFT.}
\\ [3pt]
$\bullet \;$ The second contribution
$\tfrac{\delta S[h]}{\delta h{\mu\nu}} \vert_{ind}$
represents the quantum induced stress tensor resulting 
from {\it ``integrating out''} the differentiated 
gravitons from the equations of motion in the presence 
of a constant graviton background.
\\ [3pt]
$\bullet \;$ The third contribution
$\tfrac{\delta S_{matter}}{\delta h{\mu\nu}}$ is present
in the most general case, represents classical matter 
sources, and shall not concern us further here.

A very schematical way to think of the LLOG equation
(\ref{QGrule}) is that it consists of a stochastic ``kinetic'' 
part $\tfrac{\delta S[h]}{\delta h{\mu\nu}} \vert_{stoch}$
and an induced ``potential'' part
$\tfrac{\delta S[h]}{\delta h{\mu\nu}} \vert_{ind}$ .

The gravitational interactions possess up to two 
derivatives so that at most graviton field bilinears 
will possess derivatives. The process of integrating 
out these singly or doubly differentiated graviton 
bilinears amounts to replacing them with singly or 
doubly differentiated graviton propagators in the 
presence of a constant graviton background, and this 
is tantamount to replacing them with singly or doubly
differentiated de Sitter graviton propagators with a 
different Hubble parameter.

Moreover, we must take into account that the graviton 
field contains both dynamical and constrained components: 
\begin{equation}
\kappa \, h_{\alpha\beta}
\equiv 
A_{\alpha\beta}
+ u_{(\alpha} B_{\beta)}
+ ( u_{\alpha} u_{\beta} \!+\! \overline{\gamma}_{\alpha\beta} ) C
\quad , \quad
u^{\alpha} A_{\alpha\beta} = u^{\alpha} B_{\alpha} = 0
\; , \label{ABC}
\end{equation}
where $A_{\alpha\beta}$ physically represents the dynamical 
graviton, $B_{\alpha}$ is canonically associated with the 
momentum constraints of general relativity and physically 
represents the relativistic cousins of the Newtonial potential, 
and $C$ is canonically associated with the Hamiltonian 
constraint of general relativity and physically represents 
the Newtonian potential. 

We can 3+1 decompose equations (\ref{L1+2+3eom}-\ref{Lgheom})
by contracting them alternatively with 
${\overline \gamma}_{\rho\mu} {\overline \gamma}_{\sigma\nu}$,
$u_{\mu} {\overline \gamma}_{\rho\mu}$ and, $u_{\mu} u_{\nu}$
to obtain the purely spatial, mixed spatial-temporal and,
purely temporal parts respectively.

A basic tool are the $D=4$ ghost and graviton propagators 
(\ref{gravprop},\ref{ghostprop}):
\begin{eqnarray}
i[\mbox{}_{\mu} \widetilde{\Delta}_{\rho}](x;x') 
&\!\!\! = \!\!\!&
\overline{\gamma}_{\mu\rho} \!\times i\widetilde{\Delta}_A(x;x') 
- u_{\mu} u_{\rho} \!\times i \widetilde{\Delta}_{B}(x;x')
\; , \label{ghostprop4} \\
i[\mbox{}_{\mu\nu} \widetilde{\Delta}_{\rho\sigma}](x;x') 
&\!\!\! = \!\!\!&
\sum_{I=A,B,C} [\mbox{}_{\mu\nu} \widetilde{T}^I_{~\rho\sigma}] 
\!\times i \widetilde{\Delta}_{I}(x;x') 
\; . \label{gravprop4}
\end{eqnarray}
The integrating out procedure shall require the coincidence limits 
of the three scalar propagators $\widetilde{\Delta}_{I}(x;x')$ and 
their derivatives; they are given in Table~\ref{Limits} of Appendix. 
An elementary inspection of Table~\ref{Limits} shows that the
{\it only} entry which is not finite is 
$i{\widetilde \Delta}_A (x;x') \vert_{x=x'}$ which {\it never} 
appears in any of the steps that lead to the final form of the 
LLOG equations we shall obtain. This is the justification for
computing in the $D=4$ limit.  

Furthermore, the tensor factors 
$[\mbox{}_{\mu\nu} \widetilde{T}^{I}_{~\rho\sigma}]$ 
(\ref{T_A}-\ref{T_C}) in $D=4$ are:
\begin{eqnarray}
[\mbox{}_{\mu\nu} \widetilde{T}^{A}_{~\rho\sigma}] 
&\!\!\! = \!\!\!& 
2 \, \overline{\gamma}_{\mu (\rho} \overline{\gamma}_{\sigma) \nu} 
- 2 \, \overline{\gamma}_{\mu\nu} \overline{\gamma}_{\rho\sigma} 
\; , \label{TA4} \\
{[} \mbox{}_{\mu\nu} \widetilde{T}^{B}_{~\rho\sigma}] 
&\!\!\! = \!\!\!& 
-4 \, u_{(\mu} \overline{\gamma}_{\nu) (\rho} u_{\sigma)} 
\; , \label{TB4} \\
{[} \mbox{}_{\mu\nu} \widetilde{T}^{C}_{~\rho\sigma}] 
&\!\!\! = \!\!\!&
[u_{\mu} u_{\nu} + \overline{\gamma}_{\mu\nu}] 
[u_{\rho} u_{\sigma} + \overline{\gamma}_{\rho\sigma}] 
\; , \label{TC4}
\end{eqnarray}
and their various traces -- needed for our reductions 
-- are given in Table~\ref{Traces} of the Appendix.
Additionally, a basic list of the identities used 
in deriving the LLOG gravitational equations can 
as well be found in the Appendix. In the sub-sections 
that follow we present the results of our computations.


\subsection{The``stochastic reduction'' results}

Starting from the full gravitational equations 
(\ref{L1+2+3eom}-\ref{L6eom}) we apply the relevant 
LLOG rules.
\\ [3pt]
$\bullet \;$ Concerning the 3+1 graviton field
decomposition (\ref{ABC}), the understanding is that
the purely spatial dynamical field $A_{\alpha\beta}$
is {\it stochastic} while the mixed spatial-temporal
constrained field $B_{\alpha}$ and the purely temporal 
constrained field $C$ are obviously not.
\\ [3pt]
$\bullet \;$ Concerning the 3+1 graviton field
derivatives decomposition, we have:
\begin{equation}
\kappa h_{\rho\sigma , \beta} 
= 
- u_{\beta} u \!\cdot\! \partial [ A_{\rho\sigma} 
\!-\! \mathcal{A}_{\rho\sigma} ]
+ u_{(\rho} B_{\sigma),\beta} 
+ [\overline{\gamma}_{\rho\sigma} 
\!+\! u_{\rho} u_{\sigma}] \, C_{,\beta} 
\; , \label{dABC}
\end{equation}
where $\mathcal{A}_{\rho\sigma}$ is the stochastic 
jitter free field.

Regarding the derivatives, it is worth noting that
the inflationary physical environment makes time
derivatives dominant over spatial derivatives and,
furthermore, since the evolution of the field is much
slower than that of the scale factor $a \!=\! e^{H t}$ 
the dominant contribution comes when the time 
derivative acts on the scale factor and not the 
fields. This is partly reflected in (\ref{dABC})
and partly in the following:
\\ [3pt]
{\bf I.} At each order in the dynamical $A_{\rho\sigma}$ 
field retain only the terms with  no derivatives {\it and} 
with the smallest number of derivatives,
\footnote{For instance, products of differentiated
$A_{\rho\sigma}$ fields are ignored.}
\\ 
{\bf II.} For the linear terms in $A_{\rho\sigma}$, 
each time derivative has a stochastic source
$\mathcal{A}_{\rho\sigma}$ subtracted.
\\ [3pt]
{\bf III.} All non-linear terms in the constrained 
fields $B_{\sigma}$ and $C$ are ignored. The same 
is true for terms which involve the product of a 
constrained field times a differentiated dynamical 
field $A_{\rho\sigma}$.
\footnote{The constrained fields $B_{\sigma}$ and 
$C$, as expected, are driven by the dynamical field 
$A_{\rho\sigma}$ and vanish when the latter field 
vanishes. Because $A_{\rho\sigma}$ evolves slowly we 
have schematically $B \!\sim\! C \!\sim\! \partial A
\!\sim\! small$, so that any products of them are even
smaller and can be ignored.}

Applying the above reduction rules I and II to the 
gravitational equations (\ref{L1+2+3eom}-\ref{L6eom}) 
gives:
\begin{eqnarray}
\kappa \, 
\frac{\delta S[h]_{(1+2+3)}}{\delta h_{\mu\nu}} \Big\vert_{stoch}
& \!\!\!=\!\!\! &
- \tfrac32 a^4 \sqrt{-\widetilde{g}} \times \! 
a^{-1} \widetilde{H} u \!\cdot\! \partial \Big[ 
A^{\mu\nu} \!-\! \mathcal{A}^{\mu\nu}
\!-\! \tfrac12 \widetilde{g}^{\mu\nu} ( A \!-\! \mathcal{A} ) \Big]  
\nonumber \\
& & \hspace{-1.5cm} 
+ \tfrac12 u^{(\mu} \widetilde{D}_B B^{\nu)} 
\!+\! u^{\mu} u^{\nu} \widetilde{D}_C C 
+ a^4 \sqrt{-\widetilde{g}} \, \widetilde{H}^2 \!\times\! 
u^{(\mu} \overline{\gamma}^{\nu) \alpha} A_{\alpha\beta} B^{\beta}
\; . \qquad \label{S_1+2+3_stoch}
\end{eqnarray}
\begin{eqnarray}
\kappa \,
\frac{\delta S[h]_{(4+5)}}{\delta h_{\mu\nu}} \Big\vert_{stoch}
& \!\!\!=\!\!\! &
a^4 \sqrt{-\widetilde{g}} \, \Bigl\{ 
-3 \widetilde{H}^2 u^{(\mu} \, \overline{\gamma}^{\nu) \alpha} 
A_{\alpha\beta} B^{\beta} 
\nonumber \\
& & \hspace{-2.9cm} 
- a^{-1} \widetilde{H} u^{(\mu} \, \overline{\gamma}^{\nu) \alpha} 
A_{\alpha\beta} \Big[ u \!\cdot\! \partial B^{\beta} 
\!+\! 2 \overline{\gamma}^{\beta\rho} \partial_{\rho} C \Big] 
- a^{-1} \widetilde{H} \, \widetilde{g}^{\rho (\mu} A^{\nu) \sigma} 
\partial_{\rho} B_{\sigma} \Bigr\}
\; . \qquad \label{S_4+5_stoch}
\end{eqnarray}
\begin{eqnarray}
\kappa \,
\frac{\delta S[h]_{(6)}}{\delta h_{\mu\nu}} \Big\vert_{stoch}
& \!\!\!=\!\!\! &
a^3 \sqrt{-\widetilde{g}} \, \Bigl\{
\tfrac12 \widetilde{g}^{\mu\nu} N H A^{\alpha\beta} 
\partial_{\alpha} B_{\beta} 
+ 2 H \delta^{(\mu}_{~~0} A^{\nu) \rho} \partial_{\rho} C \Bigr\}
\; . \qquad \label{S_6_stoch}
\end{eqnarray}

Next, the sum of the above contributions 
(\ref{S_1+2+3_stoch}-\ref{S_6_stoch}) is 3+1 decomposed
into purely spatial ``A'', mixed spacial-temporal ``B''
and, purely temporal ``C'' parts:
\begin{eqnarray}
\kappa \,
\frac{\delta S[h]_{A}}{\delta h_{\mu\nu}} \Big\vert_{stoch}
& \!\!\!=\!\!\! &
a^3 \sqrt{-\widetilde{g}} \, \widetilde{H} \Bigl\{
- \tfrac32 u \!\cdot\! \partial \Big[ 
A^{\mu\nu} \!-\! \mathcal{A}^{\mu\nu}
\!-\! \tfrac12 \overline{\gamma}^{\mu\nu} (A \!-\! \mathcal{A}) \Big]
\nonumber \\
& & \hspace{-1.9cm} 
- \overline{\gamma}^{\rho (\mu} A^{\nu) \sigma} \partial_{\rho} B_{\sigma} 
+ \tfrac12 N^2 \, \overline{\gamma}^{\mu\nu} A^{\alpha\beta} 
\partial_{\alpha} B_{\beta} 
+ 2 N \widetilde{g}_{0 \alpha} \overline{\gamma}^{\alpha (\mu} A^{\nu) \rho} 
\partial_{\rho} C \Bigr\}
\; , \qquad \label{S_totA_stoch}
\end{eqnarray}
\begin{eqnarray}
\kappa \,
\frac{\delta S[h]_{B}}{\delta h_{\mu\nu}} \Big\vert_{stoch}
& \!\!\!=\!\!\! &
\tfrac12 u^{(\mu} \widetilde{D}_B B^{\nu)} 
+ a^3 \sqrt{-\widetilde{g}} \, \widetilde{H} u^{(\mu} A^{\nu) \rho} \Bigl\{ 
u \!\cdot\! \partial B_{\rho} \!+\! 2N^2 \partial_{\rho} C \Bigr\} 
\nonumber \\
& & \hspace{-1.9cm}
+ a^4 \sqrt{-\widetilde{g}} \, u^{(\mu} \, \overline{\gamma}^{\nu) \alpha} 
A_{\alpha\beta} \, \Bigl\{
- \widetilde{H} a^{-1} \Big[ u \!\cdot\! \partial B^{\beta} 
\!+\! 2 \overline{\gamma}^{\beta\gamma} \partial_{\gamma} C \Big] 
-2 \widetilde{H}^2 B^{\beta} \Bigr\}
\; , \qquad \label{S_totB_stoch}
\end{eqnarray}
\begin{eqnarray}
\kappa \,
\frac{\delta S[h]_{C}}{\delta h_{\mu\nu}} \Big\vert_{stoch}
& \!\!\!=\!\!\! &
u^{\mu} u^{\nu}
\nonumber \\
& & \hspace{-0.9cm}
\times \, \Bigl\{ \widetilde{D}_C C 
+ a^3 \sqrt{-\widetilde{g}} \,
\widetilde{H} \Big[ - \tfrac34 u \!\cdot\! \partial (A \!-\! \mathcal{A}) 
- \tfrac12 N^2 A^{\alpha\beta} \partial_{\alpha} B_{\beta} \Big] \Bigr\}
\; . \qquad \label{S_totC_stoch}
\end{eqnarray}

Finally, we apply the reduction rule III which eliminates
some of the terms in (\ref{S_totA_stoch}-\ref{S_totC_stoch})
to give:
\footnote{In the case of (\ref{S_totC_stoch_fin}) we additionally 
used the ``trace identity'' (\ref{Atrace}) from section 5.2 
of the Appendix.}  
\begin{eqnarray}
\kappa \,
\frac{\delta S[h]_{A}}{\delta h_{\mu\nu}} \Big\vert_{stoch}
& \!\!\!=\!\!\! &
a^3 \sqrt{-\widetilde{g}} \, \widetilde{H} \Bigl\{
- \tfrac32 u \!\cdot\! \partial \Big[ 
A^{\mu\nu} \!-\! \mathcal{A}^{\mu\nu}
\!-\! \tfrac12 \overline{\gamma}^{\mu\nu} (A \!-\! \mathcal{A}) \Big]
\nonumber \qquad\qquad\; \\
& &
- \overline{\gamma}^{\rho (\mu} A^{\nu) \sigma} \partial_{\rho} B_{\sigma} 
+ \tfrac12 N^2 \, \overline{\gamma}^{\mu\nu} A^{\alpha\beta} 
\partial_{\alpha} B_{\beta} \Bigr\}
\; , \label{S_totA_stoch_fin}
\end{eqnarray}
\begin{eqnarray}
\kappa \,
\frac{\delta S[h]_{B}}{\delta h_{\mu\nu}} \Big\vert_{stoch}
& \!\!\!=\!\!\! &
\tfrac12 u^{(\mu} \widetilde{D}_B B^{\nu)} 
- a^3 \sqrt{-\widetilde{g}} \, \widetilde{H} u^{(\mu} A^{\nu) \rho} \Bigl\{ 
3 u \!\cdot\! \partial B_{\rho} \!+\! ( 1 \!-\! N^2) \partial_{\rho} C \Bigr\} 
\nonumber \\
& & 
+ a^4 \sqrt{-\widetilde{g}} \, u^{(\mu} A^{\nu) \rho} B_{\rho} 
\; , \qquad \label{S_totB_stoch_fin}
\end{eqnarray}
\begin{eqnarray}
\kappa \,
\frac{\delta S[h]_{C}}{\delta h_{\mu\nu}} \Big\vert_{stoch}
& \!\!\!=\!\!\! &
u^{\mu} u^{\nu}
\Bigl\{ \widetilde{D}_C C 
+ a^4 \sqrt{-\widetilde{g}} \, \tfrac{\kappa^2 \widetilde{H}^4}{8\pi^2}
\Big[ 6 - 3 {\overline \gamma}^{\alpha\beta} - 3C \Big]  
\qquad\qquad\qquad\nonumber \\
& &
- a^3 \sqrt{-\widetilde{g}} ( 1 +  N^2 )A^{\alpha\beta} 
\partial_{\alpha} B_{\beta} \Bigr\}
\; . \qquad \label{S_totC_stoch_fin}
\end{eqnarray}

\subsection{The ``integrating out'' results}

Starting from the full equations (\ref{L1+2+3eom}-\ref{Lgheom})
we must apply the relevant LLOG rules:
\\ [3pt]
{\bf IV.} Integrate out the differentiated graviton fields
from the gravitational equations (\ref{L1+2+3eom}-\ref{L6eom}).
The gravitational interactions possess up to two derivatives 
so that at most graviton field bilinears will possess 
derivatives. The process of integrating out these singly 
or doubly differentiated graviton bilinears amounts to 
replacing them with singly or doubly differentiated 
graviton propagators in the presence of a constant 
graviton background, and this is tantamount to replacing 
them with singly or doubly differentiated de Sitter 
graviton propagators with a different Hubble parameter.
\\ [3pt]
{\bf V.} Integrate out the ghost fields from the ghost
equation (\ref{Lgheom}).
\footnote{The ghost equation (\ref{Lgheom}) also
contains differentiated graviton fields. If we also
integrate out the latter, we shall produce terms
with the same logarithmic content but their overall
coefficient would then be $\kappa^4 {\widetilde H}^4$
and hence sub-leading with respect to the
$\kappa^2 {\widetilde H}^2$ coefficient produced 
by only integrating out the ghost fields.}

First, we exhibit the results for the induced stress after
applying rules IV and V. Source $S_{123}$ gives:
\begin{equation}
\frac{\delta S[h]_{(1+2+3)}}{\delta h_{\mu\nu}} \Big\vert_{ind}
= 
a^4 \sqrt{-{\widetilde g}} \, \tfrac{\kappa {\widetilde H}^4}{8\pi^2}
\Big[\! -\! \tfrac12 \overline{\gamma}^{\mu\nu} 
\!+ \tfrac{13}{2} u^{\mu} u^{\nu} \Big]
\; . \label{S_1+2+3_Tmn}
\end{equation}
The tensor structure on the RHS indicates that this source 
contributes only to the $A_{\mu\nu}$ and $C$ equations.

The induced stress from $S_{4+5}$ is:
\begin{equation}
\frac{\delta S[h]_{(4+5)}}{\delta h_{\mu\nu}} \Big\vert_{ind}
= 
a^4 \sqrt{-{\widetilde g}} \, \tfrac{\kappa^2 {\widetilde H}^4}{8\pi^2}
\, u^{\alpha} h_{\alpha\beta} 
\Big[\! - 8 \overline{\gamma}^{\beta (\mu} u^{\nu)} 
+ 12 u^{\beta} u^{\mu} u^{\nu} \Bigr]
\; . \label{S_4+5_Tmn}
\end{equation}
The tensor structure on the RHS indicates that this source 
contributes only to the $B_{\mu}$ and $C$ equations.

The induced stress from $S_{6}$ is:
\begin{eqnarray}
\frac{\delta S[h]_{(6)}}{\delta h_{\mu\nu}} \Big\vert_{ind}
& \!\!\!=\!\!\! &
a^4 \sqrt{-{\widetilde g}} \, \tfrac{\kappa {\widetilde H}^3 H}{8\pi^2}
\Bigl\{
12 N \, \widetilde{g}^{\alpha (\mu} u^{\nu)} u^{\beta} \kappa h_{\alpha\beta} 
\!-\! 12 u^{\mu} u^{\nu} u^{\beta} \kappa h_{\beta 0} 
\quad \nonumber \\
& & 
+ u^{\rho} \kappa^2 h_{\rho \alpha} h_{\beta 0} \big[ \,
6 \overline{\gamma}^{\alpha (\mu} \overline{\gamma}^{\nu) \beta} 
\!+\! 4 \overline{\gamma}^{\alpha\beta} \overline{\gamma}^{\mu\nu} 
\!+\! 2 \overline{\gamma}^{\alpha\beta} u^{\mu} u^{\nu} 
\nonumber \\
& &
- 4 u^{\alpha} \overline{\gamma}^{\beta (\mu} u^{\nu)} 
\!+\! 16 \overline{\gamma}^{\alpha (\mu} u^{\nu)} u^{\beta} 
\!-\! 24 u^{\alpha} u^{\beta} u^{\mu} u^{\nu} \, \big] \Bigr\}
\; , \label{S_6_Tmn}
\end{eqnarray}
and this source contributes to all three kinds of gravitational 
equations.

Lastly, the induced stress from $S_{gh}$ is:
\begin{eqnarray}
\frac{\delta S[h]_{gh}}{\delta h_{\mu\nu}} \Big\vert_{ind}
& \!\!\!=\!\!\! &
a^4 \sqrt{-{\widetilde g}} \, \tfrac{\kappa {\widetilde H}^4}{8\pi^2}
\Bigl\{ -\tfrac32 \overline{\gamma}^{\mu\nu} 
\!-\! \tfrac52 u^{\mu} u^{\nu} 
\!+\! \overline{\gamma}^{\mu\nu} 
[\, \overline{\gamma}^{\alpha\beta} \!+\! u^{\alpha} u^{\beta} \,] 
\kappa h_{\alpha\beta} 
\qquad \qquad \nonumber \\
& & \hspace{0.3cm}
- 5 u^{(\mu} \overline{\gamma}^{\nu) \alpha} u^{\beta} 
\kappa h_{\alpha\beta} 
\!+\! u^{\mu} u^{\nu} 
[\, \overline{\gamma}^{\alpha\beta} 
\!+\! 4 u^{\alpha} u^{\beta} \,] \kappa h_{\alpha\beta} \Bigr\} 
\; , \label{S_gh_Tmn}
\end{eqnarray}
and also contributes to all three kinds of gravitational
equations.

The {\it total} induced tensor is:
\begin{eqnarray}
\frac{\delta S[h]_{tot}}{\delta h_{\mu\nu}} \Big\vert_{ind}
& \!\!\!=\!\!\! &
a^4 \sqrt{-{\widetilde g}} \, \tfrac{\kappa {\widetilde H}^4}{8\pi^2}
\Bigl\{ \overline{\gamma}^{\mu\nu} \Big[ 
\!-\! 2 \!+\! ( \overline{\gamma}^{\alpha\beta} \!+\! u^{\alpha} u^{\beta} ) 
\kappa h_{\alpha\beta} \Big] 
\nonumber \\
& & \hspace{-2.3cm} 
- (13 \!-\! 12 N^2) u^{(\mu} \overline{\gamma}^{\nu) \alpha} 
u^{\beta} \kappa h_{\alpha\beta} 
+ u^{\mu} u^{\nu} \Big[ 4 \!+\! \Big( \overline{\gamma}^{\alpha\beta} 
\!\!+\! (16 \!-\! 12 N^2) u^{\alpha} u^{\beta} \!\Big) \kappa h_{\alpha\beta} \Big] 
\qquad \nonumber \\
& & \hspace{-2.3cm} 
- 12 N u^{\mu} u^{\nu} u^{\alpha} \kappa h_{\alpha 0} 
\!+\! N \Big[\, 6 \overline{\gamma}^{\alpha (\mu} 
\overline{\gamma}^{\nu) \beta} 
\!+\! 4 \overline{\gamma}^{\mu\nu} \overline{\gamma}^{\alpha\beta} 
\!+\! 2 u^{\mu} u^{\nu} \, \overline{\gamma}^{\alpha\beta} 
\!-\! 4 u^{(\mu} \overline{\gamma}^{\nu) \beta} u^{\alpha} 
\nonumber \\
& & \hspace{-1.3cm} 
+ 16 u^{(\mu} \overline{\gamma}^{\nu) \alpha} u^{\beta}
\!-\! 24 u^{\mu} u^{\nu} u^{\alpha} u^{\beta} \Big] 
u^{\rho} \kappa h_{\rho \alpha} \kappa h_{\beta 0} \Bigr\}
\; . \label{S_total_Tmn}
\end{eqnarray}

Next, equation (\ref{S_total_Tmn}) simplifies when 
we use (\ref{ABC}) and 3+1 decompose it into a purely 
spatial ``A'' part:
\begin{eqnarray}
\frac{\delta S[h]_{A}}{\delta h_{\mu\nu}} \Big\vert_{ind}
& \!\!\!=\!\!\! &
a^4 \sqrt{-{\widetilde g}} \, \tfrac{\kappa {\widetilde H}^4}{8\pi^2}
\Bigl\{ \overline{\gamma}^{\mu\nu} \Big[ 
\!-\! 2 \!+\! ( \overline{\gamma}^{\alpha\beta} \!+\! u^{\alpha} u^{\beta} ) 
\kappa h_{\alpha\beta} \Big] 
\nonumber \\
& & \hspace{0.3cm} 
+ N \Big[\, 6 \overline{\gamma}^{\alpha (\mu} 
\overline{\gamma}^{\nu) \beta} 
\!+\! 4 \overline{\gamma}^{\mu\nu} \overline{\gamma}^{\alpha\beta}
\Big] u^{\rho} \kappa h_{\rho \alpha} \kappa h_{\beta 0} \Bigr\}
\; , \qquad\qquad \label{TmntotA}
\end{eqnarray}
a mixed spatial-temporal ``B'' part: 
\begin{eqnarray}
\frac{\delta S[h]_{B}}{\delta h_{\mu\nu}} \Big\vert_{ind}
& \!\!\!=\!\!\! &
a^4 \sqrt{-{\widetilde g}} \, \tfrac{\kappa {\widetilde H}^4}{8\pi^2}
\nonumber \\
& &
\times \, u^{(\mu} \overline{\gamma}^{\nu) \alpha} u^{\beta} 
\Bigl\{ ( 3 \!-\! 4 N^2 ) \kappa h_{\alpha\beta} 
\!-\! 4 N \kappa h_{\beta\gamma} u^{\gamma} \kappa h_{\alpha 0} \Bigr\}
\; , \qquad\qquad \label{TmntotB}
\end{eqnarray}
and a purely temporal ``C'' part:
\begin{eqnarray}
\frac{\delta S[h]_{C}}{\delta h_{\mu\nu}} \Big\vert_{ind}
& \!\!\!=\!\!\! &
a^4 \sqrt{-{\widetilde g}} \, \tfrac{\kappa {\widetilde H}^4}{8\pi^2}
\, u^{\mu} u^{\nu}
\nonumber \\
& & \hspace{-2.3cm}
\times \, 
\Bigl\{ -8 \!+\! 12 N^2 \!+\! \Big[ \overline{\gamma}^{\alpha\beta} 
\!-\! ( 8 \!-\! 12 N^2 ) u^{\alpha} u^{\beta}\Big] \kappa h_{\alpha\beta} 
+ 2 N \, \overline{\gamma}^{\alpha\beta} u^{\rho} 
\kappa h_{\alpha\rho} \kappa h_{\beta 0} \Bigr\}
\; . \qquad \label{TmntotC}
\end{eqnarray}

Finally, application of reduction rule III further simplifies
the last set of equations (\ref{TmntotA}-\ref{TmntotC}) to a
very reasonable form:
\begin{equation}
\frac{\delta S[h]_{A}}{\delta h_{\mu\nu}} \Big\vert_{ind}
=
a^4 \sqrt{-{\widetilde g}} \, \tfrac{\kappa {\widetilde H}^4}{8\pi^2}
\Bigl\{ \overline{\gamma}^{\mu\nu} \Big[ 
\!-\! 2 \!+\! ( \overline{\gamma}^{\alpha\beta} \!+\! u^{\alpha} u^{\beta} ) 
\kappa h_{\alpha\beta} \Big] \Bigr\}
\; , \qquad\qquad\qquad \label{TmntotA2}
\end{equation}
\begin{equation}
\frac{\delta S[h]_{B}}{\delta h_{\mu\nu}} \Big\vert_{ind}
=
- a^4 \sqrt{-{\widetilde g}} \, \tfrac{\kappa {\widetilde H}^4}{16\pi^2}
( 3 \!-\! 4 N^2 ) u^{(\mu} B^{\nu)}
\; , \qquad\qquad\qquad\qquad\qquad\quad \label{TmntotB2}
\end{equation}
\begin{equation}
\frac{\delta S[h]_{C}}{\delta h_{\mu\nu}} \Big\vert_{ind}
=
a^4 \sqrt{-{\widetilde g}} \, \tfrac{\kappa {\widetilde H}^4}{8\pi^2}
\, u^{\mu} u^{\nu}
\Bigl\{ -8 \!+\! 12 N^2 
\!+\! \overline{\gamma}^{\alpha\beta} A_{\alpha\beta}
\!-\! ( 8 \!-\! 12 N^2 ) C \Bigr\}
\; . \;\; \label{TmntotC2}
\end{equation}


\subsection{The final results}

In view of the basic equation (\ref{QGrule}):
\begin{equation} 
\kappa \frac{\delta S[h]_I}{\delta h_{\mu\nu}} \Big\vert_{stoch}
= 
- \kappa \frac{\delta S[h]_I}{\delta h_{\mu\nu}} \Big\vert_{ind}
\quad , \quad I = A, B, C
\; , \label{Ifinal}
\end{equation}
and the partial results from sub-sections 3.2 \& 3.3, 
the full LLOG equations immediately follow in their most 
optimal form:
\footnote{Equation (\ref{Afinal}) follows from 
(\ref{S_totA_stoch_fin},\ref{TmntotA2}), 
equation (\ref{Bfinal}) from (\ref{S_totB_stoch_fin},\ref{TmntotB2}) 
after eliminating the common factor $u^{\mu}$, 
and equation (\ref{Cfinal}) from (\ref{S_totC_stoch_fin},\ref{TmntotC2})
after eliminating the common factor $u^{\mu} u^{\nu}$.}
\begin{eqnarray}
u \!\cdot\! \partial (A^{\mu\nu} \!-\! \mathcal{A}^{\mu\nu}) 
& \!\!\!=\!\!\! & 
\tfrac43 \frac{\kappa^2 \widetilde{H}^3}{8 \pi^2} 
a \, \overline{\gamma}^{\mu\nu} 
\Big[ 2 \!-\! \overline{\gamma}^{\alpha\beta} A_{\alpha\beta} \!-\! 4C \Big] 
\nonumber \\
& &  
- \tfrac23 \overline{\gamma}^{\rho (\mu} A^{\nu) \sigma} 
\partial_{\rho} B_{\sigma} 
\!+\! \tfrac23 ( 1 \!-\! N^2 ) \overline{\gamma}^{\mu\nu} A^{\alpha\beta} 
\partial_{\alpha} B_{\beta} 
\; , \qquad\qquad\qquad \label{Afinal}
\end{eqnarray}
\begin{eqnarray}
\widetilde{D}_B B^{\mu} 
& \!\!\!=\!\!\! &
\frac{\kappa^2 \widetilde{H}^4}{8 \pi^2} 
a^4 \sqrt{-\widetilde{g}} \, (3 \!-\! 4 N^2) B^{\mu}  
\nonumber \\
& & 
+ a^3 \sqrt{-\widetilde{g}} \, \widetilde{H} A^{\mu\nu} 
\Bigl\{ 4( 1 \!-\! N^2 ) \partial_{\nu} C 
\!+\! 4 a \widetilde{H} B_{\nu} \Bigr\} 
\; , \qquad\qquad\quad\; \label{Bfinal} 
\end{eqnarray}
\begin{eqnarray}
\widetilde{D}_C C 
& \!\!\!=\!\!\! & 
\frac{\kappa^2 \widetilde{H}^4}{8 \pi^2} 
a^4 \sqrt{-\widetilde{g}} \, \Bigl\{ 
14 \!-\! 12 N^2 \!-\! 4 \overline{\gamma}^{\alpha\beta} A_{\alpha\beta} 
\!-\! (7 \!+\! 12 N^2) C \Bigr\} 
\qquad\qquad\qquad \nonumber \\
& & 
+ a^3 \sqrt{-\widetilde{g}} \, \widetilde{H} 
(1 \!-\! N^2) A^{\alpha\beta} \partial_{\alpha} B_{\beta} 
\; . \label{Cfinal}
\end{eqnarray}

The field variables are $A_{ij}, B_i, C$ and are related 
to the geometrical variables $\gamma_{ij}, N^i, N$ thusly:
\footnote{Equations (\ref{N}-\ref{gamma_ij}) can be 
straightforwardly derived by expressing 
$d{\widetilde s}^{\,\,\! 2} = 
{\widetilde g}_{\mu\nu} dx^{\mu} dx^{\nu}$ in two different 
ways: $d{\widetilde s}^{\,\,\! 2} = 
(\eta_{\mu\nu} + \kappa h_{\mu\nu}) dx^{\mu} dx^{\nu}$ with 
$h_{\mu\nu}$ given by equation (\ref{ABC}) and, \\
$d{\widetilde s}^{\,\,\! 2} = - N^2 d\eta^2 + \gamma_{ij} 
(dx^i - N^i d\eta) (dx^j - N^j d\eta)$. Thereafter, we use 
identities from sub-section 5.2 of the Appendix.}
\begin{eqnarray}
N^2 
& \!\!\!=\!\!\! &
\frac{1}{1 + C + \tfrac14 B_i B_i}
\; , \label{N} \\
N^i 
& \!\!\!=\!\!\! &
\frac{\tfrac12 B_i}{\sqrt{1 + C + \tfrac14 B_i B_i}}
\; , \label{Ni} \\
\gamma_{ij}
& \!\!\!=\!\!\! &
\frac{\delta_{ij} + A_{ij}}{1 - C}
\; , \label{gamma_ij} 
\end{eqnarray}

Motivated by relations (\ref{N}-\ref{gamma_ij}) and 
the slow growth of the constrained fields $B^i$ and 
$C$, we can further simplify (\ref{Afinal}-\ref{Cfinal})
using:
\begin{equation}
N^2 \sim 1-C
\quad , \quad 
u \!\cdot\! \partial \sim a \, \partial_t
\quad , \quad
\widetilde{D}_J J \sim 
-2 {\widetilde H}^2 a ^4 \sqrt{-\widetilde{g}} \, J 
\,\, {\scriptstyle (J = B^i , C)}
\; . \label{simplify2}
\end{equation}

The result of these simplifications is a remarkably 
simple form for the gauge fixed LLOG gravitational
operator equations in accelerating background spacetimes:
\begin{eqnarray}
{\dot A}^{\mu\nu} \!-\! {\dot {\mathcal A}}^{\mu\nu} 
& \!\!\!=\!\!\! & 
\tfrac43 \frac{\kappa^2 \widetilde{H}^3}{8 \pi^2} 
\overline{\gamma}^{\mu\nu} 
\Big[ 2 \!-\! \overline{\gamma}^{\alpha\beta} A_{\alpha\beta} \!-\! 4C \Big] 
- \tfrac{2}{3} \, a^{-1} \, \overline{\gamma}^{\rho (\mu} A^{\nu) \sigma} 
\partial_{\rho} B_{\sigma} 
\; , \qquad \label{Afinal2} \\
B^{\mu} 
& \!\!\!=\!\!\! &
\frac{\kappa^2 \widetilde{H}^2}{16 \pi^2} B^{\mu}  
- 2 A^{\mu\nu} B_{\nu} 
\; , \label{Bfinal2} \\
C 
& \!\!\!=\!\!\! & 
\frac{\kappa^2 \widetilde{H}^2}{8 \pi^2} 
\Big[ - 1 \!+\! 2 \overline{\gamma}^{\alpha\beta} A_{\alpha\beta} 
\!+\! \tfrac72 C \Bigr] 
\; . \label{Cfinal2}
\end{eqnarray}

\section{Epilogue}

This work represents the final step in the long sequence of
generalizations of Starobinsky's stochastic formalism 
\cite{Starobinsky:1986fx} from scalar potential models 
\cite{Tsamis:2005hd} to interactions with passive fields
\cite{Miao:2006pn,Prokopec:2007ak}, non-linear sigma models
\cite{Miao:2021gic}, and finally MMC scalar corrections to
gravity \cite{Miao:2024nsz}. The new features for pure gravity
are tensor indices, which required the new gauge developed in
the previous paper \cite{Miao:2024shs}, and constrained fields,
which formed an important part of the analysis in the current 
paper. A crucial aspect of all previous steps 
\cite{Tsamis:2005hd,Miao:2006pn,Prokopec:2007ak,Miao:2021gic,
Miao:2024nsz} in this program was the comparison of stochastic
predictions with explicit, dimensionally regulated and fully
renormalized computations at 1-loop and 2-loop orders. That
has not yet be done for our formulation of stochastic quantum
gravity, and we enjoin caution until this painstaking process
has been completed.

The methodology used to derive the LLOG eqs 
(\ref{Afinal2}-\ref{Cfinal2}) is a realization 
of the underlying physical mechanism whereby 
real particle production in the inflationary 
background -- a quantum process -- induces a 
stress tensor and thereupon gravity responds 
classically to its presence \cite{Tsamis:2014kda}.
This is the physical interpretation of the two terms 
on the RHS of the ``rule equation" (\ref{QGrule}); 
the stochastic term represents the leading logarithm 
classical gravitational response to the quantum 
induced stress tensor term.

The set of Feynman rules presented not only allows 
explicit invariantly regularized and renormalized 
computations in perturbation theory, but also allows
the derivation of the operator field equations 
appropriate for re-sum\-ming the leading logarithms. 
Regarding the latter, it should be apparent it is 
a well defined procedure because the dimensionless 
coupling constant of the theory $G \Lambda \ln(a)$ 
contains two parameters and hence the classification 
of diagrams by their $\ln(a)$ dependence is well 
defined.

Furthermore, expressions (\ref{Afinal2}-\ref{Cfinal2}) 
are gauge fixed operator equations rather than
effective field equations. Therefore, if we insert the
operator solution of the equations into an observable 
and compute its vacuum expectation value, we should 
obtain gauge independent results. Such a phenomenologically
relevant procedure could be executed for a variety of 
observables, for instance:
\\ [3pt]
{\it (i)} the expansion rate, 
\\ [3pt]
{\it (ii)} the gravitational force due to a test mass,
\\ [3pt]
{\it (iii)} the effective gravitational theory emerging
after the accelerating era,
\\ [3pt]
{\it (iv)} the graviton mode function and the resulting
tensor primordial power spectrum.

In concluding, perhaps we should re-iterate that the 
operator equations (\ref{Afinal2}-\ref{Cfinal2}) turned 
out to have a surprisingly simple form. These equations 
contain the leading contribution to all the physical 
effects associated with real particle production in 
accelerating spacetimes. It is apparent that an important
application would be to compute the predicted values 
of any of the above observables by considering the 
vacuum expectation value of the corresponding operator. 
Indeed this is work in progress \cite{pathIII}. 

\vskip 0.5cm

\centerline{\bf Acknowledgements}

This work was partially supported by Taiwan NSTC grants 
113-2112-M-006-013 and 114-2112-M-006-020, by NSF grant 
PHY-2207514 and by the Institute for Fundamental Theory 
at the University of Florida.


\newpage

\section{Appendix: Useful Identities \& Tables}

\subsection{Some relations from the 3+1 decomposition:}

\begin{eqnarray}
& \mbox{} &
{\widetilde H} = \frac{H}{N}
\quad , \quad 
{\widetilde H}^2 = \frac{H^2}{N^2} 
= - {\widetilde g}^{00} H^2 
\; , \label{Htilde} \\
& \mbox{} &
\delta^0_{\, \mu} = - \frac{1}{N} u_{\mu}
\quad , \quad
{\widetilde g}^{0\mu} = - \frac{1}{N} u^{\mu}
\quad , \quad
{\widetilde g}^{\mu\nu} u_{\mu} u_{\nu} 
= -1 =
{\widetilde g}_{\mu\nu} u^{\mu} u^{\nu} 
\; . \qquad\qquad \label{u's}
\end{eqnarray}

\subsection{Some tensor algebra identities:}

\begin{equation}
{\widetilde g}^{\mu\nu} =
\eta^{\mu\nu} - \kappa h^{\mu}_{\; \alpha} \; {\widetilde g}^{\alpha\nu}
\; , \label{ID1}
\end{equation}
\begin{equation}
\eta_{\mu\nu} =
{\overline \eta}_{\mu\nu} - \delta^0_{\; \mu} \delta^0_{\; \nu}
\quad , \quad
\delta^{\mu}_{\; \nu} =
{\overline \delta}^{\mu}_{\nu} + \delta^{\mu}_{\; 0} \delta^0_{\; \nu}
\; , \label{ID2}
\end{equation}
\begin{equation}
\delta^{\mu}_{\; \alpha} \delta^{\nu}_{\; \beta}
=
{\overline \delta}^{\mu}_{\; (\alpha} 
{\overline \delta}^{\nu}_{\; \beta)}
+ 2 \delta^{(\mu}_{\;\; 0} \, {\overline \delta}^{\nu)}_{\; (\alpha}
  \delta^0_{\; \beta)}
+ \delta^{\mu}_{\; 0} \delta^{\nu}_{\; 0}
  \delta^0_{\; \alpha} \delta^0_{\; \beta}
\; , \label{ID3}
\end{equation}
\begin{equation}
{\widetilde g}^{\mu\nu}_{~~ , \alpha} =
- {\widetilde g}^{\mu\rho} \, {\widetilde g}^{\nu\sigma}
\kappa h_{\rho\sigma , \alpha}
\quad , \quad
\delta^{\mu}_{\; 0} =
{\widetilde g}^{\mu\rho} \, {\widetilde g}_{\rho 0} =
{\overline \gamma}^{\mu\rho} \, {\widetilde g}_{\rho 0}
- u^{\mu} u_0
\; , \label{ID4}
\end{equation}
\begin{equation}
\kappa h_{\mu 0} =
{\widetilde g}_{\mu 0} - \eta_{\mu 0} =
{\widetilde g}_{\mu 0} + \delta^0_{\; \mu} =
\tfrac12 u_0 B_{\mu} + u_0 u_{\mu} C 
\; , \label{ID5}
\end{equation}
\begin{equation}
u^{\alpha} \kappa h_{\alpha\beta} =
- \tfrac12 B_{\beta} - u_{\beta} C
\quad , \quad
u^{\alpha} u^{\beta} \kappa h_{\alpha\beta} =
C
\; , \label{ID6}
\end{equation}
\begin{equation}
{\overline \gamma}^{\alpha\beta} \kappa h_{\alpha\beta} =
{\overline \gamma}^{\alpha\beta} A_{\alpha\beta}
\quad , \quad 
{\overline \gamma}^{\mu \alpha} A_{\alpha\beta} =
\Big[ {\widetilde g}^{\mu\alpha} + u^{\mu} u^{\alpha} \Big] 
A_{\alpha\beta} =
A^{\mu}_{\; \beta}
\; . \label{ID7}
\end{equation}
\begin{equation}
u \cdot \partial (A \!-\! \mathcal{A}) = 
\tfrac{\kappa^2 {\widetilde H}^3}{2 \pi^2} a \Big[ 
2 \!-\! \overline{\gamma}^{\alpha\beta} A_{\alpha\beta} \!-\! 4 C \Big] 
+ (\tfrac43 \!-\! 2 N^2) A^{\alpha\beta} \partial_{\alpha} B_{\beta} 
\; . \label{Atrace}
\end{equation}

\subsection{Quadratic operator properties:} 

$\bullet \,$ 
The ${\widetilde D}_A$ quadratic operator equals:
\begin{equation}
{\widetilde D}_A 
\equiv
\partial_{\alpha} \big[ a^{D-2} {\sqrt {-\widetilde g}} \,
{\widetilde g}^{\alpha\beta} \partial_{\beta} \big]
\; , \label{D_A}
\end{equation}
and its operation on the three kinds of scalar propagators
gives:
\begin{eqnarray}
{\widetilde D}_A \, i {\widetilde \Delta}_A(x;x') 
&\!\!\!=\!\!\!& 
i \delta^D(x-x')
\; , \label{iDeltaA} \\
{\widetilde D}_A \, i {\widetilde \Delta}_B(x;x')
&\!\!\!=\!\!\!& 
i \delta^D(x-x') 
+ (D-2) {\widetilde H}^2 a^D {\sqrt {-\widetilde g}} \;
  i {\widetilde \Delta}_B(x;x')
\; , \label{iDeltaB} \\
{\widetilde D}_A \, i {\widetilde \Delta}_C(x;x')
&\!\!\!=\!\!\!& 
i \delta^D(x-x') 
+ 2(D-3) {\widetilde H}^2 a^D {\sqrt {-\widetilde g}} \; 
  i {\widetilde \Delta}_C(x;x') 
\; . \qquad \label{iDeltaC}
\end{eqnarray}
$\bullet \,$
The other two quadratic operators equal:
\begin{eqnarray}
{\widetilde D}_B 
&\!\!\!=\!\!\!&
{\widetilde D}_A - (D-2) \widetilde{H}^2 a^D \sqrt{-\widetilde{g}}
\; , \label{D_B} \\
{\widetilde D}_C 
&\!\!\!=\!\!\!& 
\widetilde{D}_A - 2(D-3) \widetilde{H}^2 a^D \sqrt{-\widetilde{g}}
\; . \label{D_C}
\end{eqnarray}
In $D \!=\! 4$ we have $(D-2) = 2(D-3) = 2$ so that 
${\widetilde D}_B = {\widetilde D}_C$ .

\newpage

\noindent
$\bullet \,$
The graviton quadratic operator equals: 
\begin{equation}
\widetilde{\mathcal D}^{\mu\nu\alpha\beta}
=
\tfrac12 \Big[ {\widetilde g}^{\mu(\alpha} \, {\widetilde g}^{\beta)\nu}
  - \tfrac12 {\widetilde g}^{\mu\nu} \, {\widetilde g}^{\alpha\beta} \Big] 
  \widetilde{\mathcal D}_A
+ (D-2){\widetilde H}^2 a^D \sqrt{- {\widetilde g}} \,
  u^{(\mu} \, {\widetilde g}^{\nu)(\alpha} \, u^{\beta)}
\; , \label{Dgrav}
\end{equation}
and allows us to check that indeed the proper condition
is satisfied:
\begin{equation}
\widetilde{\mathcal D}^{\mu\nu\alpha\beta} \, 
i \Bigl[ \mbox{}_{\alpha\beta} \widetilde{\Delta}_{\rho\sigma} \Bigr](x;x') 
=
\delta^{\mu}_{\; (\rho} \delta^{\nu}_{\; \sigma)} \,
i \delta^D (x-x')
\; . \label{propcheck}
\end{equation}

\noindent
$\bullet \,$
The ghost kinetic operator equals:
\begin{equation}
\widetilde{\mathcal D}^{\mu\alpha}
=
{\widetilde g}^{\mu\alpha} {\widetilde D}_A 
+ (D-2) {\widetilde H}^2 a^D {\sqrt {-\widetilde g}} \,
u^{\mu} u^{\alpha} 
\; , \label{ghostquadratic} 
\end{equation}
so that its action on (\ref{ghostprop}) gives:
\begin{equation}
\widetilde{\mathcal D}^{\mu\alpha} \, 
i \Bigl[ \mbox{}_{\alpha} \widetilde{\Delta}_{\rho} \Bigr](x;x') 
=
\delta^{\mu}_{\; \rho} \, i \delta^D (x-x')
\; . \label{ghostpropcheck}
\end{equation}

\subsection{Propagator coincidence limits in $D=4$ dimensions:}
\begin{table}[H]
\setlength{\tabcolsep}{8pt}
\def\arraystretch{1.5}
\centering
\begin{tabular}{|@{\hskip 1mm }c@{\hskip 1mm }||c|c|c|}
\hline
{\rm Limit} & $i\widetilde{\Delta}_{A}$ & $i\widetilde{\Delta}_{B}$ & 
$i\widetilde{\Delta}_{C}$ \\
\hline\hline
$i\widetilde{\Delta}_{I}(x;x')\vert_{x'=x}$ & 
$\tfrac{{\widetilde H}^2}{4\pi^2} \ln a + ``\infty"$ &
$-\tfrac{{\widetilde H}^2}{16\pi^2}$ & 
$+\tfrac{{\widetilde H}^2}{16\pi^2}$ \\
\hline
$\partial_{\mu} i\widetilde{\Delta}_{I}(x;x')\vert_{x'=x}$ & 
$-\tfrac{{\widetilde H}^3}{8\pi^2} \, a u_{\mu}$ & 
$0$ & 
$0$ \\
\hline
$\partial_{\mu} \partial_{\nu} i\widetilde{\Delta}_{I}(x;x')\vert_{x'=x}$ 
& $\tfrac{{\widetilde H}^4}{8\pi^2} \, a^2 
[ \frac34 \widetilde{g}_{\mu\nu} + u_{\mu} u_{\nu} ]$ 
& $-\tfrac{{\widetilde H}^4}{32\pi^2} \, a^2 \widetilde{g}_{\mu\nu}$ 
& $+\tfrac{{\widetilde H}^4}{32\pi^2} \, a^2 \widetilde{g}_{\mu\nu}$ \\
\hline
$\partial_{\mu} \partial'_{\nu} i\widetilde{\Delta}_{I}(x;x')\vert_{x'=x}$ 
& $-\tfrac{{3\widetilde H}^4}{32\pi^2} \, a^2 \widetilde{g}_{\mu\nu}$ 
& $+\tfrac{{\widetilde H}^4}{32\pi^2} \, a^2 \widetilde{g}_{\mu\nu}$ 
& $-\tfrac{{\widetilde H}^4}{32\pi^2} \, a^2 \widetilde{g}_{\mu\nu}$ \\
\hline
\end{tabular}
\caption{\footnotesize Coincidence limits of propagators 
and their derivatives in $D=4$ dimensions. In the above 
relations, we first take the derivatives and then the 
coincidence limit $x' \rightarrow x$.} 
\label{Limits}
\end{table}

\subsection{Tensor factor traces in $D=4$ dimensions:}
\begin{table}[H]
\setlength{\tabcolsep}{8pt}
\def\arraystretch{1.5}
\centering
\begin{tabular}{|@{\hskip 1mm }c@{\hskip 1mm }||c|c|c|c|}
\hline
{\rm Tensor\ Factor} & $[\mbox{}_{\mu\nu} \widetilde{T}^{\rho}_{~\rho}]$ 
& $[\mbox{}^{\rho}_{~\rho} \widetilde{T}^{\sigma}_{~\sigma}]$ 
& $[\mbox{}^{\rho}_{~\nu} \widetilde{T}_{\rho \sigma}]$ 
& $[\mbox{}^{\rho\sigma} \widetilde{T}_{\rho\sigma}]$ \\
\hline\hline
$[\mbox{}_{\mu\nu} \widetilde{T}^{A}_{~\rho\sigma}]$ 
& $-4 \overline{\gamma}_{\mu\nu}$ & $-12$ 
& $+2\overline{\gamma}_{\nu\sigma}$ & $+6$ \\
\hline
$[\mbox{}_{\mu\nu} \widetilde{T}^{B}_{~\rho\sigma}]$ 
& $0$ & $0$ & $+\overline{\gamma}_{\nu\sigma} - 3 u_{\nu} u_{\sigma}$ & $+6$ \\
\hline
$[\mbox{}_{\mu\nu} \widetilde{T}^{C}_{~\rho\sigma}]$ 
& $\!\!+2 \overline{\gamma}_{\mu\nu} \!+\! 2 u_{\mu} u_{\nu}\!\!$ & $+4$ 
& $\!\!+\overline{\gamma}_{\nu\sigma} \!-\! u_{\nu} u_{\sigma}\!\!$ & $+4$ \\
\hline
$\frac34 [\widetilde{T}^{A}] - \frac14 [\widetilde{T}^{B}] 
+ \frac14 [\widetilde{T}^{C}]$
& $\!\!-\frac52 \overline{\gamma}_{\mu\nu} \!+\! 
\frac12 u_{\mu} u_{\nu}\!\!$ & $-8$ 
& $\!\!+\frac32 \overline{\gamma}_{\nu\sigma} \!+\! 
\frac12 u_{\nu} u_{\sigma}\!\!$ & $+4$ \\
\hline
\end{tabular}
\caption{\footnotesize Traces of tensor factors in $D=4$ dimensions.}
\label{Traces}
\end{table}

\newpage


\end{document}